# Total Reflection and Negative Refraction of Dipole-Exchange Spin Waves at Magnetic Interfaces: Micromagnetic Modeling Study

Sangkook Choi and Sang-Koog Kim*

Research Center for Spin Dynamics & Spin-Wave Devices, Seoul National University, Seoul 151-744, Republic of Korea
Nanospintronics Laboratory, Department of Materials Science and Engineering, College of Engineering, Seoul National University, Seoul 151-744, Republic of Korea

We demonstrated that dipole-exchange spin waves traveling in geometrically restricted magnetic thin films satisfy the same laws of reflection and refraction as light waves. Moreover, we found for the first time novel wave behaviors of dipole-exchange spin waves such as total reflection and negative refraction. The total reflection in laterally inhomogeneous thin films composed of two different magnetic materials is associated with the forbidden modes of refracted dipole-exchange spin waves. The negative refraction occurs at a 90° domain-wall magnetic interface that is introduced by a cubic magnetic anisotropy in the media, through the anisotropic dispersion of dipole-exchange spin waves.

Dipole-exchange spin waves (DESWs) in ordered magnetic materials of restricted geometry have attracted considerable interest in the research fields of magnetization (**M**) dynamics [1]. DESWs are low-lying collective excitations of **M**s, and are mediated by both long-range dipolar and short-range exchange interactions in submicron sized magnetic elements. Along with DESW eigenmodes in patterned magnetic thin films [2], DESWs in the gigahertz frequency range have attracted much attention, due to their potential uses not only as signals in new-paradigm logic devices [3] but also as monochromatic "on board" gigahertz-frequency-range microwave sources in integrated electronic circuits [4]. In order to realize such promising DESW applications, it is necessary to steer the path of DESW rays propagating in two-dimensional (2D) magnetic media with high-energy flux [5]. In the field of wave optics, such fundamental and technological issues in 2D or 3D optical media have been described well on the basis of the optical laws of reflection and refraction at various interfaces. Owing to the difference in phase velocity of light waves between two dissimilar media, which is also macroscopically described as the difference between the refractive indices of dissimilar media, the path of wave rays can be controlled using optical components such as a mirror and a prism. In particular, the novel wave behaviors of total reflection and negative refraction enable us to guide the energy flux of wave rays with negligible energy losses and to realize superlens, superprism and open-cavity of light waves [6], respectively. In an analogous manner, we expect that the propagation direction of DESWs can also be controlled through such reflection and refraction phenomena. However, these wave properties at magnetic interfaces have been studied only for spin waves mediated by either exchange interaction or magnetic dipolar interaction, not for spin waves mediated by both interactions in restricted geometry [7].

In this Letter, we demonstrate, numerically, that DESWs satisfy the same laws of reflection and refraction as light waves. Based on those wave phenomena, we elucidate for the first time the total reflection and negative refraction effects of DESWs in laterally inhomogeneous ferromagnetic thin films and their underlying physics. The results might lead the way in realizing potential information-processing devices based on the wave properties of DESWs propagating in magnetic waveguide circuits.

Before we proceed further into the present study, first let us consider the optical laws of reflection and refraction with light waves of a certain frequency $f$ incident on an interface between media 1 and 2, not only in real ($x$ - $y$) space but also in wavevector ($k_x$ - $k_y$) space, as illustrated respectively in Figs. 1(a) and 1(b) [8]. When plane waves of a wavevector $\mathbf{k^0}$ and a group velocity $\mathbf{v_g^0}$ are incident on an interface parallel to the $x$-axis, some of them return to (are reflected into) medium 1 with $\mathbf{k'}$ and $\mathbf{v_g'}$, and the rest of them pass through the interface (are refracted) and continue to propagate in medium 2 with $\mathbf{k''}$ and $\mathbf{v_g''}$. The general relations between the incident, reflected, and refracted waves are as follows; $k_x^0 = k_x' = k_x''$ at the interface, and the signs of $v'_{g,y}$ and $v''_{g,y}$ are opposite to and the same as that of $v^0_{g,y}$, where $v_{g,y} = \mathbf{v_g} \cdot \hat{e}_y$ with $\mathbf{v_g} \equiv \nabla_\mathbf{k} f^{medium}(\mathbf{k})$, as shown in Fig. 1(b).

To examine whether DESWs obey the same laws of reflection and refraction as light waves, we conducted micromagnetic simulations [9] on the incidence of DESWs with $f = 51$ GHz at an incidence angle of 45° from Ni (y < 0 nm) [10] to Permalloy ($Ni_{80}Fe_{20}$: Py, y > 0 nm) [11] media of 5 nm thickness, as shown in Fig. 2(a). These two media were saturated along the $+y$ direction by an externally applied field of 1 Tesla. To produce a monochromatic DESW ($f = 51$ GHz) with flat wavefront normal to the 45° incidence angle, an oscillating magnetic field $H_{osc} = 250 (1 - \cos2\pi vt)$ Oe was applied along the $+x$ axis only in the dash-dot boxed region, where $v = 51$ GHz [5]. Figure 2(b) shows a snapshot image of the spatial distribution, in the area marked by the gray-colored dotted box in Fig. 2(a), of the out-of-plane components, $M_z$, of the local **M**s normalized by the saturation **M**, $M_s$, taken at the time, $t = 4$ ns. When the propagating DESWs encountered the Ni / Py interface, some of them were reflected, thus forming a distinct interference pattern by the superposition of the incident and reflected DESWs in the Ni medium, and the remainder continued to propagate in the Py medium through the interface, showing the refraction characteristics. For further quantitative analysis, we made Fast-Fourier-transform (FFT) of the spatial distribution of the local $M_z / M_s$, as shown in Fig. 2(c). There were three strong peaks on the vertical line placed at $k_x = 0.09$ nm$^{-1}$, positioned at $(k_x, k_y) = (0.09$ nm$^{-1}$, 0.09

nm$^{-1}$), (0.09 nm$^{-1}$, 0.06 nm$^{-1}$), and (0.09 nm$^{-1}$, -0.09 nm$^{-1}$). These three peaks were identified, according to the frequency contours of the allowed DESW modes for $f$ = 51 GHz in the Ni (purple) and Py (apricot) media, as the individual incident (red), reflected (blue), and refracted (green) DESWs [12]. The incident, reflected, and refracted DESWs had the same value of $k_x$, reflecting the fact that DESWs obey the same laws as do light waves in reflection and refraction.

On the basis of the reflection and refraction behaviors of DESWs confirmed above, we demonstrated the total reflection of DESWs occurring in a certain incidence angle range. In the case of light waves, the energy flux of incident waves is totally reflected if the allowed modes of $|k_x''|$ in medium 2, equal to those of $|k_x^0|$ in medium 1, do not exist. To demonstrate the novel behavior of DESWs at certain incidence angles, we considered the same model geometry under the same external static field as in Fig. 2(a), but the Ni and Py were replaced with Py and $Y_3Fe_5O_{12}$ (YIG) [13] materials. To excite the DESWs with an $f$ = 60 GHz of flat wavefronts normal to the incidence angle of 45°, $H_{osc}$ with $v$ = 60 GHz was applied along the + $x$ axis. Figure 3(a) shows the resultant spatial distribution of the $M_z / M_s$ components taken, upon applying $H_{osc}$, at $t$ = 4.0 ns. When the DESWs were incident at the interface, they were reflected, thus forming a distinct interference pattern in the Py medium. By contrast, there were no refracted DESWs in the YIG medium. For a more quantitative analysis, we obtained the FFT power distribution of the local $M_z / M_s$, finding two strong peaks at $(k_x, k_y)$= (0.11 nm$^{-1}$, 0.11 nm$^{-}$

$^{-1}$) and (0.11 nm$^{-1}$, -0.11 nm$^{-1}$), as seen in Fig. 3(b). The purple-colored line, which represents the frequency contour of DESW modes with $f$ = 60 GHz in the Py medium, allows us to identify the incident (red) and reflected (blue) DESWs [12], in accordance with the reflection law ($k_x^0 = k_x'$). Under this condition, there were no DESW modes available in the YIG medium, as evidenced by the absence of any excitation peak on the apricot-colored line corresponding to the allowed modes of DESWs in the YIG medium.

From the frequency contours for $f$ = 60 GHz in the Py and YIG thin films shown in Fig. 3(c), we can estimate the incidence-angle ranges, marked by the gray-colored regions, in which the total reflection of DESWs is available at the Py / YIG interface, according to the condition under which the values of $|k_x^0|$ in the Py medium are larger than the maximum value of $|k_x''|$ = 0.05 nm$^{-1}$ in the YIG medium. In the angle regions ranging from 0° to 72.8° and from 107.2° to 180°, the condition of $k_x'' = k_x^0$ required for refraction of the incident DESWs is no longer satisfied. The total reflection for the 45° incidence angle shown in Figs. 3(a) and 3(b) is one among those available angles.

We also report on an intriguing negative refraction behavior of DESWs. Negative refraction at an interface along the $x$-axis in optics indicates that the sign of $v_{g,x}''$ is opposite to that of $v_{g,x}^0$, which can be possible either at an interface between a normal right-handed material and a left-handed material of negative permittivity and permeability [14], or at an interface made by an anisotropic medium [15]. Let us consider the latter case in order to demonstrate the negative refraction of DESWs. In a medium with anisotropic dispersion, the direction of $\mathbf{v_g}$ (gray colored thick arrows) can differ from that of $\mathbf{k}$ (gray colored thin arrows), as illustrated in Fig. 4(a). Consequently, $v_{g,x}$ and $k_x$ in the anisotropic medium can have either the same or the opposite sign, according to the direction of $\mathbf{k}$ relative to the direction of the optical axis of the medium, as shown in Fig. 4(a). When $v_{g,x}$ and $k_x$ have the opposite sign in medium 1 and the same sign in medium 2, $v_{g,x}^0$ and $v_{g,x}''$ can be of the opposite sign owing to the law of refraction, which makes a negative refraction by definition possible, as visualized in Fig. 4(b).

In a similar way, DESWs also exhibit an anisotropic dispersion relation: the magnitude of $\mathbf{k}$ varies with its direction with respect to the orientation of the static $\mathbf{M}$. Analogously to the case of light waves, DESWs are also expected to be negatively refracted under peculiar conditions. The following is our numerical demonstration using a model system of 5 nm-thick Fe media with a cubic magnetic anisotropy [16]. The cubic anisotropy saturates the static magnetization orientations along the direction $(x, y, z) = (-1,1,0)$ in the $Fe_1$ region ($y < 0$ nm) and along $(x, y, z) = (1,1,0)$ in the $Fe_2$ region ($y > 0$ nm), resulting in a peculiar 90° domain-wall magnetic interface along the $x$-axis. The thorny shape at the edges of such a finite-size model system was designed to align the local magnetizations along the desired directions through the cubic anisotropy in both the $Fe_1$ and $Fe_2$ media [17]. To excite DESWs of $f = 25$

GHz incident normally on the 90° domain wall with flat wavefronts parallel to the $Fe_1$ / $Fe_2$ interface, we applied $H_{osc}$ with $\nu = 25$ GHz only in the dash-dot boxed region along the direction perpendicular to the static magnetization, as shown in Fig. 5(a).

The resultant distribution of the dynamic $M_z$ / $M_s$ components in the gray-colored dotted box in Fig. 5(a), taken at $t = 4.0$ ns after applying $H_{osc}$, is shown in Fig. 5(b). Interestingly, the DESW packets with wave front parallel to the $Fe_1$ / $Fe_2$ interface propagated at certain angles from the interface in both the $Fe_1$ and $Fe_2$ media. This clearly exhibits not only that the direction of $\mathbf{v_g}$ differs from the $\mathbf{k}$ direction for both incident and refracted DESW rays, but also that the tangential components of the propagation directions of the DESW packets to the $Fe_1$ / $Fe_2$ interface in the $Fe_1$ and $Fe_2$ media have the opposite sign. This DESW propagation is similar to the negative refraction behavior of light waves. For further quantitative analysis, we made an FFT of the local $M_z$ / $M_s$ distribution to obtain their power spectrum, as shown in Fig. 5(c). In the distribution on the $k_x$ - $k_y$ plane, there was only one strong peak at $(k_x, k_y) = (0$ nm$^{-1}$, $0.07$ nm$^{-1})$. On the FFT power distribution, $\mathbf{k}$ and $\mathbf{v_g}$ for both the incident and refracted DESWs could be identified by overlapping the DESW frequency contours with $f = 25$ GHz, which are available in the $Fe_1$ (purple) and $Fe_2$ (apricot) media [12]. At the $(k_x, k_y) = (0$ nm$^{-1}$, $0.07$ nm$^{-1})$ peak, the directions of $\mathbf{v_g}$ for the incident and refracted DESW rays were identified as 52° in the $Fe_1$ and 128° in the $Fe_2$ media respectively, from the $+ x$ direction, as determined by the definition of $\mathbf{v_g} \equiv \nabla_{\mathbf{k}} f^{Fe}(\mathbf{k})$ (see Fig. 5(c)) [12]. These angles are in quite good agreement with the directions of the DESW packets represented by the $M_z / M_s$ distribution in Fig. 5(b). The range of the incidence angles where the negative refraction of DESWs occurs can be predicted from the frequency contours of the DESW modes. For example, Figure 5(d) shows the $f$ contours for DESW modes with $f = 25$ GHz in 5 nm-thick Fe$_1$ and Fe$_2$ thin films and the signs of $v_{g,x}^0$ and $v_{g,x}''$. By the definition of $\mathbf{v_g} \equiv \nabla_{\mathbf{k}} f^{Fe}(\mathbf{k})$, the angles for $v_{g,x}^0 = 0$ and $v_{g,x}'' = 0$ were determined to be 116.1° and 63.9° from the $+x$ direction, respectively. Between these two angles, the directions of $v_{g,x}^0$ and $v_{g,x}''$ lie opposite to each other, so that the negative refraction of DESWs can take place in any angles of incidence ranging from 63.9° to 116.1°.

In conclusion, we demonstrated, by micromagnetic simulations, that DESWs in 2D magnetic media satisfy the same laws of reflection and refraction as light waves. In particular, total reflection and negative refraction of DESWs were found at various magnetic interfaces for the first time in 2D model systems. The total reflection results from the forbidden refraction mode of $k_x''$, which should be the same as $k_x^0$. Also, the negative refraction was demonstrated using the anisotropic DESW dispersion. These results would provide a basic mechanism to guide DESWs in potential information-processing devices.

This work was supported by Creative Research Initiatives (the ReC-SDSW) of MOST/KOSEF.

**Figure captions**

FIG. 1 (color online) Schematic illustrations of the reflection and refraction of light waves at an optical interface between two dissimilar media in real (*x-y*) space and wave vector ($k_x$-$k_y$) space, as shown in (a) and (b), respectively. The purple- and apricot-colored circles on the $k_x$–$k_y$ plane indicate the individual frequency contours of the allowed optical modes in the same-color-coded media 1 and 2, respectively. The color-coded thin and thick arrows denote the wave vectors **k** and group velocities **v**$_g$ of waves propagating in each medium, as indicated by the incident (red), reflected (blue), and refracted (green) rays.

FIG. 2 (color online) (a) 5 nm-thick-film model system composed of different magnetic elements of Ni (y < 0 nm) and Py (y > 0 nm). **M**s in the two different media are saturated along the + *y* direction, as noted, by applying an external magnetic field of 1 Tesla. (b) Perspective image of the spatial distribution of the $M_z$ / $M_s$ components, taken at *t* = 4 ns. (c) FFT power distribution on the $k_x$-$k_y$ space, obtained from the $M_z$ / $M_s$ distribution on the *x-y* plane. The purple- and apricot-colored curved lines represent the frequency contours of the DESW modes in the Ni and Py media, respectively.

FIG. 3 (color online) Spatial distribution of the local $M_z$ / $M_s$ components taken at *t* = 4 ns for the temporal evolution of total reflection behavior in (a) and the FFT power distribution on the $k_x$-$k_y$ space in (b). (c) The anticipated range of the incidence angles (gray-colored region) where the total reflections possibly take place. The purple- and apricot-colored curves represent the frequency contours of the DESWs in the Py and YIG media, respectively.

FIG. 4 (color online) (a) The frequency contour (apricot-colored line) of allowed optical modes in an anisotropic medium. The gray-colored thin and thick arrows denote examples of the allowed **k** and **v**$_g$ of waves propagating in the medium. The purple and yellowish green arrows represent the signs of $k_x$ and $v_{g,x}$ respectively. (b) Schematic diagram for the negative refraction of light waves at an interface between media 1 and 2 where $v_{g,x}$ and $k_x$ have the opposite sign (medium 1) and the same sign (medium 2), respectively. The purple- and apricot-colored lines represent the frequency contours of the allowed modes in media 1 and 2, respectively.

FIG. 5 (color online) (a) 5 nm-thick Fe model system for the demonstration of a DESW's negative refraction. The **M**s in each medium are saturated along the direction of $(x, y, z) = (-1,1,0)$ and $(1,1,0)$ in the Fe$_1$ (y < 0 nm) and Fe$_2$ (y > 0 nm) media, respectively, as illustrated by the colors shown on the colored wheel. (b) Plane-view image of the spatial distribution of the local $M_z / M_s$ component, taken at $t$ = 4 ns. (c) The FFT power distribution of the local $M_z / M_s$ on the $k_x$-$k_y$ plane. The purple- and apricot-colored lines represent the frequency contours of the DESWs of $f$ = 25 GHz in the Fe$_1$ and Fe$_2$ media, respectively. (d) The anticipated range of the incidence angles (gray-colored region) where the negative refractions take place.

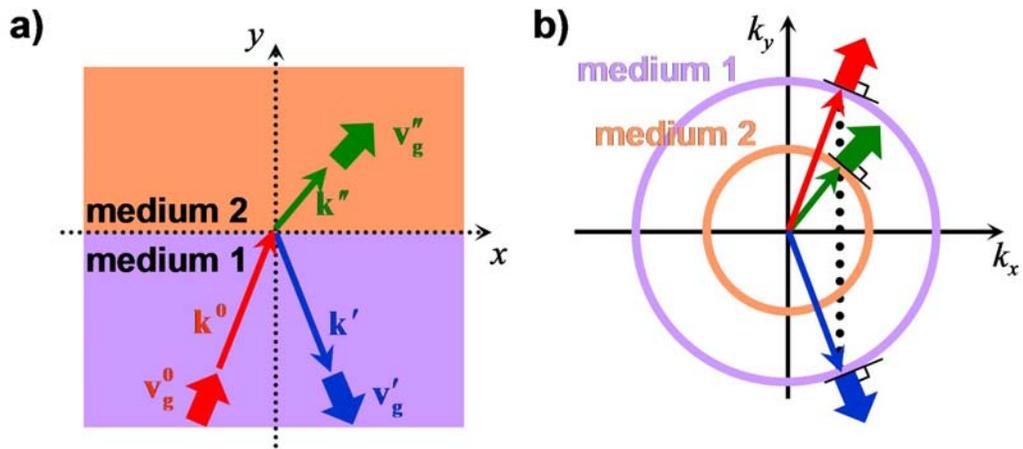

Fig. 1

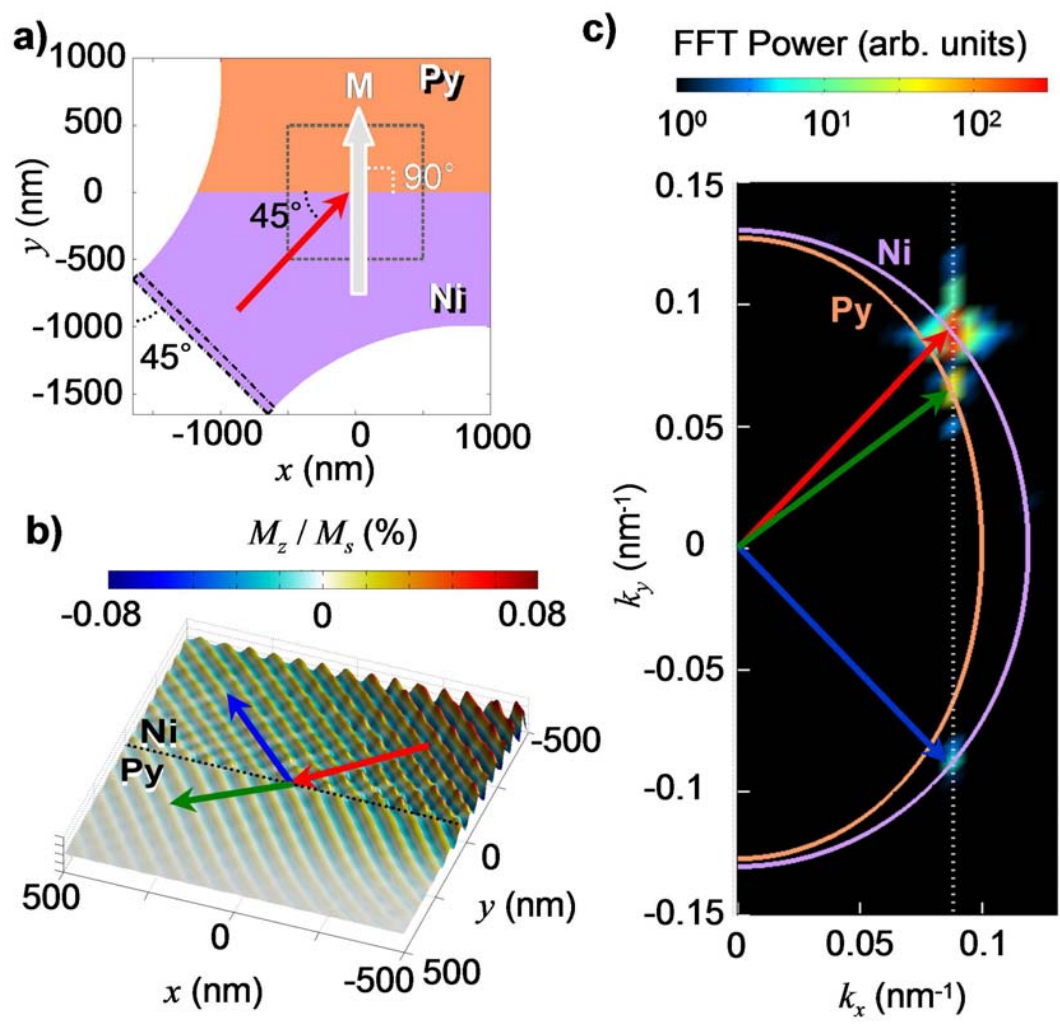

Fig. 2

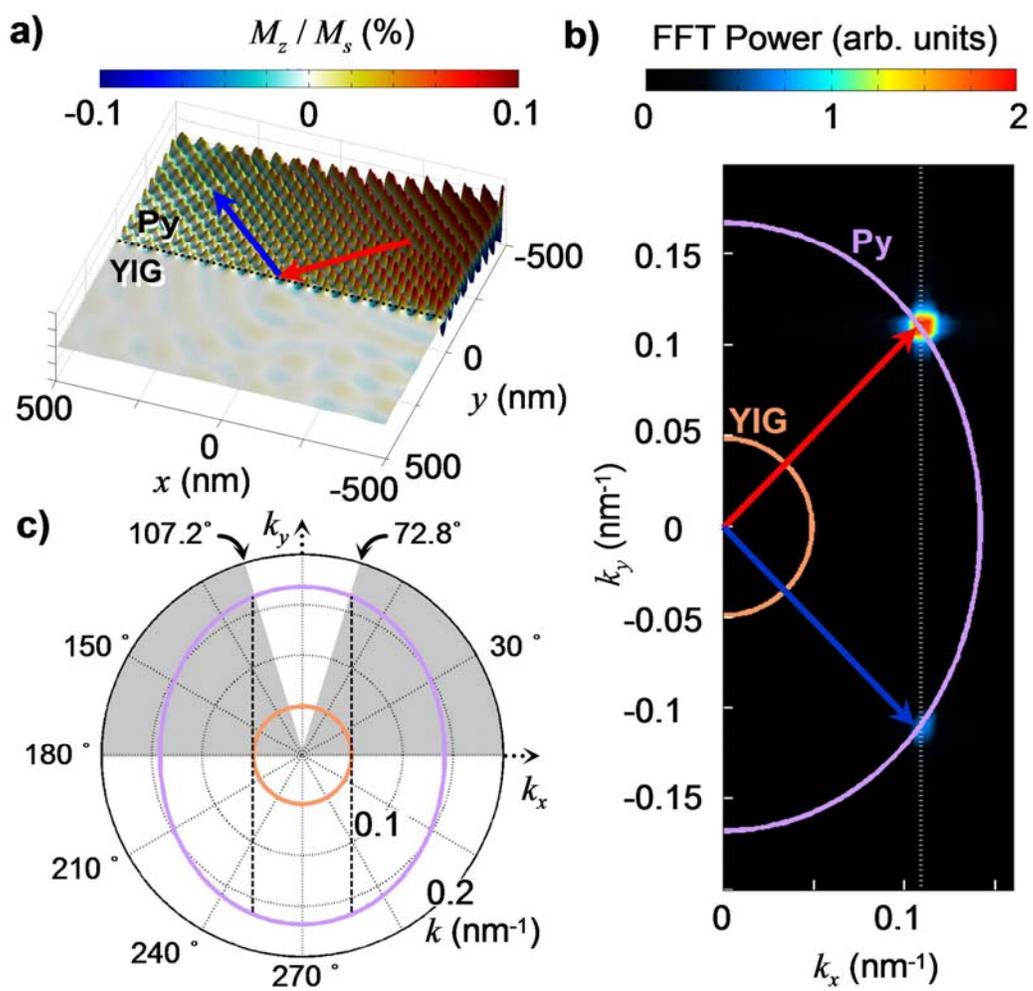

Fig.3

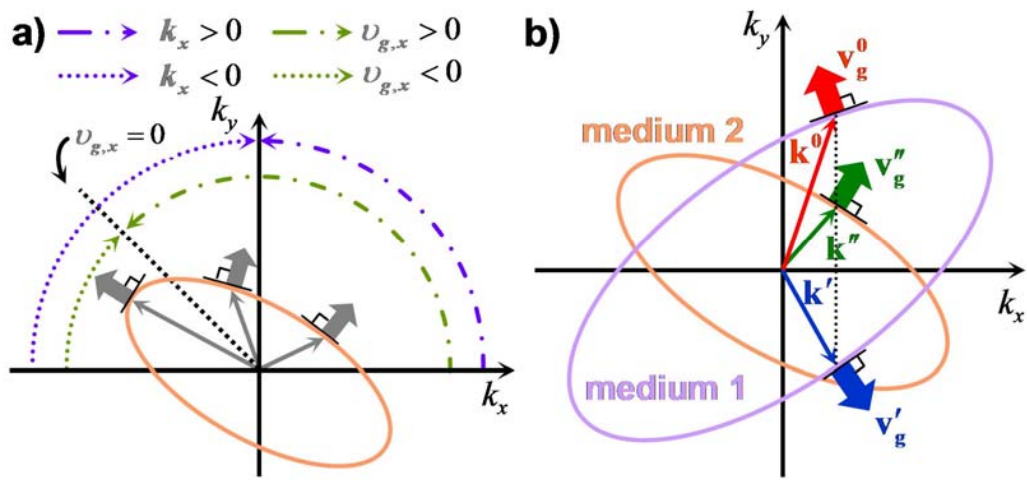

Fig. 4

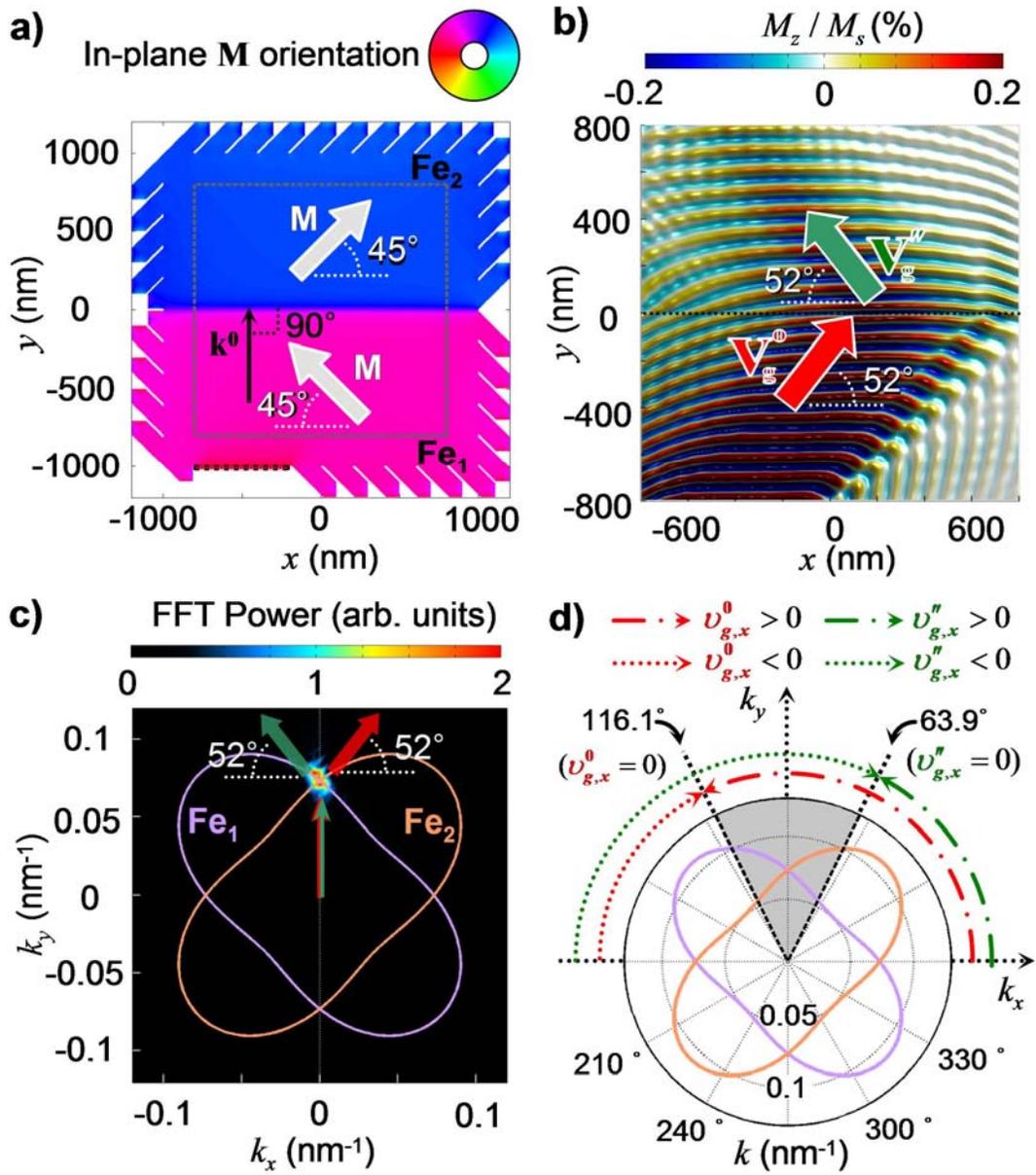

Fig. 5